\documentclass[thmsa,letterpaper,12pt]{article}
\usepackage{sw20lart}

\input tcilatex
\QQQ{Language}{
American English
}

\begin{document}

\author{Nick Laskin\thanks{%
E-mail: laskin@cims.nyu.edu}}
\title{\textbf{L\'evy Flights over Quantum Paths}\\
}
\date{New York University\\
Courant Institute of Mathematical Sciences\\
251 Mercer Street, NY 10012\\
}
\maketitle

\begin{abstract}
An impact of integration over the paths of the L\'evy flights on the quantum
mechanical kernel has been studied. Analytical expression for a free
particle kernel has been obtained in terms of the Fox $H$-function. A new
equation for the kernel of a partical in the box has been found. New general
results include the well known quantum formulae for a free particle kernel
and particle in box kernel.

\textit{PACS }number(s): 03.65.-w, 03.65. Db, 05.30.-d, 05.40. Fb

\textit{Keywords}: L\'evy flights, path integral, quantum mechanical kernel.
\end{abstract}

\section{Introduction}

In past decade it has been realized that the understanding of complex
quantum and classical physic phenomena has required the implementation of
the L\'evy flights random process \cite{Zaslavsky1}, \cite{Metzler}. It is
well known that a position of diffusive particle increases as square root in
time, $x(t)\sim t^{1/2}$. The square root law is attribute of the Brownian
motion model for diffusion. However for complex quantum and classical physic
phenomena this temporal diffusive behavior has not been observed. Instead,
more general evolution law $x(t)\sim t^{1/\alpha }$ with $0<\alpha \leq 2$
has been found. Thus, the well known diffusion law $x(t)\sim t^{1/2}$ is
included as a special case at $\alpha =2$. The mathematical model to
describe $1/\alpha $ diffusion scaling is known as L\'evy flights. The
L\'evy flights is general framework to study 'unusual diffusion' in terms of
fractional Fokker-Planck equation \cite{Zaslavsky1}, \cite{Metzler}.

The scaling $1/\alpha $ law assigned to the L\'evy flights has been
empirically observed in laser cooling of atoms \cite{Bardou}, in ion
dynamics in optical lattice \cite{Schlipf}, in anomalous transport \cite
{Zimbardo}, in the measurement of the momentum of cold cesium atoms in a
periodically pulsed standing wave of light \cite{Raizen}. The L\'evy flights
are widely used to model a variety of physical phenomena such as kinetics
and transport in classical systems, anomalous diffusion, chaotic dynamics,
plasma physics, dynamics of economic indexes, biology and physiology, social
science (see for example, \cite{Zaslavsky1}, \cite{Metzler} and references
there).

So too, the Feynman path integral ought to be generalized to describe
complex dynamic phenomena at the quantum level. The Feynman path integral is
in fact the integration over Brownian-like quantum mechanical paths \cite
{Feynman1}, \cite{Feynman2}. To generalize the Feynman path integral
approach the integration has been expanded from Brownian-like to L\'evy
flights paths \cite{Laskin1}, \cite{Laskin2}.

In this paper we study an impact of integration over the paths of the L\'evy
flights on the structure of a free particle quantum mechanical kernel. The
kernel has been expressed in terms of Fox's $H$-function.

The paper is organized as follows. We introduce path integral over the
L\'evy paths in Sec.2. It has been shown in Sec.2 that the Levy path
integral goes over into the Feynman path integral when self-similarity (or
the L\'evy) index $\alpha $=2. Thus, our new results go over into the well
known quantum equations at the special case $\alpha =2$.

Sec.3 deals with a free particle quantum kernel. We have found Fox's $H$%
-function representation for a free particle quantum kernel. The
representation gives us an option to use many well established equations,
identities, integrals involving the $H$-function for practical L\'evy path
integral calculations.

Sec.4 summarizes the Laplace and the energy-time transforms and the momentum
representation of the L\'evy quantum mechanical kernel.

New equation for the quantum kernel of a particle in a box has been found
and discussed in Sec.5.

\section{Path Integral}

\subsection{Feynman path integral}

If a particle at an initial time $t_a$ starts from the point $x_a$ and goes
to a final point $x_b$ at time $t_b$, we will say simply that the particle
goes from $a$ to $b$ and its trajectory (path) $x(t)$ will have the property
that $x(t_a)=x_a$ and $x(t_b)=x_b$. In quantum mechanics, then, we will have
an quantum-mechanical kernel, which we may write as $K(x_bt_b|x_at_a)$, to
get from the point $a$ to the point $b$. This will be the sum over all of
the trajectories that go between that end points and of a contribution from
each. Following Feynman \cite{Feynman1} we write the kernel in the form

\begin{equation}
K(x_bt_b|x_at_a)=\int\limits_{x(t_a)=x_a}^{x(t_b)=x_b}\emph{D}x(\tau )\exp
\{\frac i\hbar S(x)\},  \label{eq1}
\end{equation}

where $S(x)$ is classical mechanical action considered as the functional of
a particle trajectory $x(\tau )$

\begin{equation}
S(\tau )=\int\limits_{t_a}^{t_b}d\tau \left( \frac{m\stackrel{.}{x}^2}%
2-V(x(\tau ),\tau ))\right) ,  \label{eq2}
\end{equation}

here $V(x,t)$ is the potential energy and $\int%
\limits_{x(t_a)=x_a}^{x(t_b)=x_b}\emph{D}x(\tau )...$ is defined as follows

\begin{equation}
\int\limits_{x(t_a)=x_a}^{x(t_b)=x_b}\emph{D}x(\tau )...=\stackunder{%
N\rightarrow \infty }{\lim }\left( \frac{2\pi i\hbar \varepsilon }m\right)
^{-1/2}\int\limits_{-\infty }^\infty \prod\limits_{j=1}^{N-1}\left( \frac{%
2\pi i\hbar \varepsilon }m\right) ^{-1/2}dx_j...,\quad \varepsilon =\frac{%
t_b-t_a}N.  \label{eq3}
\end{equation}

Feynman's original path integral approach to non-relativistic quantum
mechanics is based on the fundamental equations (\ref{eq1})-(\ref{eq3}).

Then the equation

\begin{equation}
\psi (x_b,t_b)=\int\limits_{-\infty }^\infty dx_aK(x_bt_b|x_at_a)\psi
(x_a,t_a).  \label{eq4}
\end{equation}

gives the wave function $\psi (x_b,t_b)$ at a time $t_b$ in terms of the
wave function $\psi (x_a,t_a)$ at a time $t_a$. This fundamental equation
describes the evolution of the quantum mechanical system in the terms of the
wave function.

\subsubsection{The free particle}

To calculate the quantum-mechanical amplitude $K^{(0)}(x_bt_b|x_at_a)$ for a
free particle, $V(x)=0$ we will follow Feynman \cite{Feynman1} ,

\begin{equation}
K^{(0)}(x_bt_b|x_at_a)=\int\limits_{x(t_a)=x_a}^{x(t_b)=x_b}\emph{D}x(\tau
)\exp \left\{ \frac i\hbar \int\limits_{t_a}^{t_b}d\tau \frac{m\stackrel{.}{x%
}^2}2\right\} .  \label{eq5}
\end{equation}

With the help of Eqs.(\ref{eq1})-(\ref{eq3}) we have

\begin{equation}
K^{(0)}(x_bt_b|x_at_a)=\stackunder{N\rightarrow \infty }{\lim }\left( \frac{%
2\pi i\hbar \varepsilon }m\right) ^{-1/2}\int\limits_{-\infty }^\infty
\prod\limits_{j=1}^{N-1}\left( \frac{2\pi i\hbar \varepsilon }m\right)
^{-1/2}dx_j\times  \label{eq6}
\end{equation}

\[
\prod\limits_{j=1}^N\exp \left\{ \frac{im}{2\hbar \varepsilon }%
(x_j-x_{j-1})^2\right\} . 
\]

The calculation is carried out as follows. It is easy to see that \footnote{%
Here we use the definite integral
\par
\[
\int\limits_{-\infty }^\infty dxe^{-ax^2+bx}=\sqrt{\frac \pi a}e^{b^2/4a}. 
\]
}

\[
\left( \frac{2\pi i\hbar \varepsilon }m\right) ^{-1/2}\int\limits_{-\infty
}^\infty dx_1\exp \left\{ \frac{im}{2\hbar \varepsilon }\left[
(x_2-x_1)^2+(x_1-x_a)^2\right] \right\} = 
\]

\begin{equation}
\left( \frac{2\pi i\hbar 2\varepsilon }m\right) ^{-1/2}\exp \left\{ \frac{im%
}{2\hbar 2\varepsilon }(x_2-x_a)^2\right\} .  \label{eq7}
\end{equation}

Next we multiply this result by

\[
\left( \frac{2\pi i\hbar \varepsilon }m\right) ^{-1/2}\exp \left\{ \frac{im}{%
2\hbar \varepsilon }(x_3-x_2)^2\right\} , 
\]

and integrate again, this time over $x_2.$ The result of integration is
similar to that of Eq.(\ref{eq7}), except that $(x_2-x_a)^2$ becomes $%
(x_3-x_a)^2$ and the expression $2\varepsilon $ is replaced by $3\varepsilon 
$ in two places. Thus, we get

\[
\left( \frac{2\pi i\hbar 3\varepsilon }m\right) ^{-1/2}\exp \left\{ \frac{im%
}{2\hbar 3\varepsilon }(x_3-x_a)^2\right\} . 
\]

In this way a recursion procedure is established which after $N-1$ steps
gives

\[
\left( \frac{2\pi i\hbar N\varepsilon }m\right) ^{-1/2}\exp \left\{ \frac{im%
}{2\hbar N\varepsilon }(x_a-x_0)^2\right\} . 
\]

Since $N\varepsilon =t_b-t_a$ , it is easy to see that the final result
after $N-1$ steps is

\begin{equation}
K^{(0)}(x_bt_b|x_at_a)=\left( \frac{2\pi i\hbar (t_b-t_a)}m\right)
^{-1/2}\cdot \exp \left\{ \frac{im(x_b-x_a)^2}{2\hbar (t_b-t_a)}\right\} .
\label{eq8}
\end{equation}

The Eq.(\ref{eq8}) implies

\[
\Delta x\propto \left( \frac \hbar m\right) ^{1/2}(\Delta t)^{1/2}, 
\]

which means that space displacement $\Delta x=x_b-x_a$ and the time $\Delta
t=t_b-t_a$ scale are governed by the Brownian motion law.

In terms of the Fourier integral the Feynman kernel $K^{(0)}(x_bt_b|x_at_a)$
given by Eq$.(\ref{eq8})$ can be written as

\begin{equation}
K^{(0)}(x_bt_b|x_at_a)=\frac 1{2\pi \hbar }\int\limits_{-\infty }^\infty
dp\cdot \exp \left\{ i\frac{p(x_b-x_a)}\hbar -i\frac{p^2(t_b-t_a)}{2m\hbar }%
\right\} .  \label{eq9}
\end{equation}

\subsection{The path integral over the L\'evy paths}

The ''natural'' way to generalize Eq.(\ref{eq9}) is \cite{Laskin1}, \cite
{Laskin2}

\begin{equation}
K_L^{(0)}(x_bt_b|x_at_a)=\frac 1{2\pi \hbar }\int\limits_{-\infty }^\infty
dp\cdot \exp \left\{ i\frac{p(x_b-x_a)}\hbar -i\frac{D_\alpha |p|^\alpha
(t_b-t_a)}\hbar \right\} ,  \label{eq10}
\end{equation}

where $D_\alpha $ is ''fractional quantum diffusion coefficient'' physical
dimension of which is

\[
\lbrack D_\alpha ]=\mathrm{erg}^{1-\alpha }\cdot \mathrm{cm}^\alpha \cdot 
\mathrm{sec}^{-\alpha }, 
\]

and $\alpha $ is the Levy index, and we consider $1<\alpha \leq 2$.

It is easy to see from Eq.(\ref{eq10}) that the scaling relation between
space displacement $\Delta x=x_b-x_a$ and the time increment $\Delta
t=t_b-t_a$ has a form

\[
\Delta x\propto \left( \hbar ^{\alpha -1}D_\alpha \right) ^{1/\alpha
}(\Delta t)^{1/\alpha }. 
\]

This scaling relation implies that space displacement and the time increment
are governed by the L\'evy flights law.

Using Eq.(\ref{eq10}) we can define the path integral in the phase space
representation by the following way

\begin{equation}
\int\limits_{x(t_a)=x_a}^{x(t_b)=x_b}\mathrm{D}x(\tau )\int\limits_{}^{}%
\mathrm{D}p(\tau )...=  \label{eq11}
\end{equation}

\[
=\stackunder{N\rightarrow \infty }{\lim }\int\limits_{-\infty }^\infty
dx_1...dx_{N-1}\frac 1{(2\pi \hbar )^N}\int\limits_{-\infty }^\infty
dp_1...dp_N\times 
\]

\[
\exp \left\{ i\frac{p_1(x_1-x_a)}\hbar -i\frac{D_\alpha \varepsilon
|p_1|^\alpha }\hbar \right\} \times ...\times \exp \left\{ i\frac{%
p_N(x_b-x_{N-1})}\hbar -i\frac{D_\alpha \varepsilon |p_N|^\alpha }\hbar
\right\} ..., 
\]

here again $\varepsilon =(t_b-t_a)/N$. Then the kernel $K_L(x_bt_b|x_at_a)$
defined by Eq.(\ref{eq10}) can be written as

\[
K_L(x_bt_b|x_at_a)=\stackunder{N\rightarrow \infty }{\lim }%
\int\limits_{-\infty }^\infty dx_1...dx_{N-1}\frac 1{(2\pi \hbar
)^N}\int\limits_{-\infty }^\infty dp_1...dp_N\times 
\]

\[
\exp \left\{ \frac i\hbar \sum\limits_{j=1}^Np_j(x_j-x_{j-1})\right\} \times
\exp \left\{ -\frac i\hbar D_\alpha \varepsilon
\sum\limits_{j=1}^N|p_j|^\alpha -\frac i\hbar \varepsilon
\sum\limits_{j=1}^NV(x_j,j\varepsilon )\right\} . 
\]

In the continuum limit $N\rightarrow \infty ,\quad \varepsilon \rightarrow 0$
we have

\begin{equation}
K_L(x_bt_b|x_at_a)=  \label{eq12}
\end{equation}

\[
\int\limits_{x(t_a)=x_a}^{x(t_b)=x_b}\mathrm{D}x(\tau )\int\limits_{}^{}%
\mathrm{D}p(\tau )\exp \left\{ \frac i\hbar \int\limits_{t_a}^{t_b}d\tau
[p(\tau )\stackrel{\cdot }{x}(\tau )-H_\alpha (p(\tau ),x(\tau ),\tau
]\right\} , 
\]

where the phase space path integral $\int\limits_{x(t_a)=x_a}^{x(t_b)=x_b}%
\mathrm{D}x(\tau )\int\limits_{}^{}\mathrm{D}p(\tau )...$ is given by Eq.(%
\ref{eq11}), $\stackrel{\cdot }{x}$ denotes the time derivative, $H_\alpha $
is the Hamiltonian

\begin{equation}
H_\alpha (p,x)=D_\alpha |p|^\alpha +V(x,t)  \label{eq13}
\end{equation}

with the replacement $p\rightarrow p(\tau )$, $x\rightarrow x(\tau )$ and $%
\{p(\tau ),x(\tau )\}$ is the particle trajectory in phase space. The
hermiticity property of the Hamiltonian $H_\alpha $ has been discussed in 
\cite{Laskin3}.

The exponential in Eq.(\ref{eq12}) can be written as $\exp \{\frac i\hbar
S_\alpha (p,x)\}$ if we introduce canonical action $S_\alpha (p,x)$ for the
trajectory $p(t)$, $x(t)$ in phase space

\begin{equation}
S_\alpha (p,x)=\int\limits_{t_a}^{t_b}d\tau (p(\tau )\stackrel{\cdot }{x}%
(\tau )-H_\alpha (p(\tau ),x(\tau ),\tau )).  \label{eq14}
\end{equation}

Since the coordinates $x_0$, $x_N$ in the definition (\ref{eq11}) are fixed
at their initial and final points, $x_0=x_a$ and $x_N=x_b$, the all possible
trajectories in Eqs.(\ref{eq11}), (\ref{eq12}) satisfy the boundary
condition $x(t_b)=x_b$, $x(t_a)=x_a$. We see that the definition given by
Eq.(\ref{eq11}) includes one more $p_j$-integrals than $x_j$-integrals.
Indeed, while $x_0$ and $x_N$ are held fixed and the $x_j$-integrals are
done for $j=1,...,N-1$, each increment $x_j-x_{j-1}$ is accompanied by one $%
p_j$-integral for $j=1,...,N$. The above observed asymmetry is a consequence
of the particular boundary condition. Namely, the end points are fixed in
the position (coordinate) space. There exists the possibility of proceeding
in a conjugate way keeping the initial $p_a$ and final $p_b$ momenta fixed.
The associated kernel can be derived going through the same steps as before
but working in the momentum representation (see, for example, \cite{Kleinert}%
).

Taking into account Eq.(\ref{eq10}) it is easily to check on directly the
consistency condition

\[
K_L^{(0)}(x_bt_b|x_at_a)=\int\limits_{-\infty }^\infty dx^{\prime
}K_L^{(0)}(x_bt_b|x^{\prime }t^{\prime })\cdot K_L^{(0)}(x^{\prime
}t^{\prime }|x_at_a). 
\]

This is a special case of the general quantum-mechanical rule: the kernels
for events occurring in succession in time satisfy

\begin{equation}
K_L(x_bt_b|x_at_a)=\int\limits_{-\infty }^\infty dx^{\prime
}K_L(x_bt_b|x^{\prime }t^{\prime })\cdot K_L(x^{\prime }t^{\prime }|x_at_a),
\label{eq15}
\end{equation}

where $K_L(x_bt_b|x^{\prime }t^{\prime })$ is defined by Eq.(\ref{eq12}).

\section{Fox $H$ function representation for a free particle kernel $%
K_L^{(0)}(x_bt_b|x_at_a)$}

Let us show how a free particle quantum mechanical kernel $%
K_L^{(0)}(x_bt_b|x_at_a)$ defined by Eq.(\ref{eq10}) can be expressed in the
terms of the Fox's $H$-function \cite{Fox}, \cite{Mathai}, \cite{Srivastava}%
. Apart from the quiet natural way in which the Fox's $H$-function enters in
the path integral technique over the L\'evy flights, its derivatives and
integrals are easily calculated by formally manipulating the parameters in
the $H$-function. That is why the $H$-Fox's function is so important and
useful for the path integration over the L\'evy paths. Note that $H$%
-function bears the name of their discoverer Fox \cite{Fox} although it has
been known at least since 1888 (see \cite{Mathai}).

To present a free particle kernel $K_L^{(0)}(x_bt_b|x_at_a)$ in terms of the
Fox $H$-function we accept the following strategy. Starting from the
representation (\ref{eq10}) we obtain the Mellin transform of the quantum
mechanical kernel. Comparing of the inverse Mellin transform with the
definition of the Fox's function \cite{Fox}, \cite{Mathai} yields the
desired expression in terms of ''known'' function, i.e. Fox's $H$-function.

Introducing for simplicity the notations

\begin{equation}
x\equiv x_b-x_a,\qquad \tau \equiv t_b-t_a,  \label{eq16}
\end{equation}

we rewrite Eq.(\ref{eq10})

\begin{equation}
K_L^{(0)}(x,\tau )=\frac 1{2\pi \hbar }\int\limits_{-\infty }^\infty dp\cdot
\exp \left\{ i\frac{px}\hbar -i\frac{D_\alpha |p|^\alpha \tau }\hbar
\right\} .  \label{eq17}
\end{equation}

One can see that the relation

\[
K_L^{(0)}(x,\tau )=K_L^{(0)}(-x,\tau ), 
\]

holds. Hence, it is sufficient to consider $K_L^{(0)}(x,\tau )$ for $x\geq 0$
only.

Further, we will use the following definitions of the Mellin

\begin{equation}
\stackrel{\wedge }{K}(s,\tau )=\int\limits_0^\infty dxx^{s-1}K_L(x,\tau ),
\label{eq18}
\end{equation}

and the inverse Mellin transforms

\begin{equation}
K_L(x,\tau )=\frac 1{2\pi i}\int\limits_{c-i\infty }^{c+i\infty }dsx^{-s}%
\stackrel{\wedge }{K}(s,\tau ),  \label{eq19}
\end{equation}

where the integration path is the straight line from $c-i\infty $ to $%
c+i\infty $ with $0<c<1$.

The Mellin transform of the $K_L^{(0)}(x,\tau )$ is

\[
\stackrel{\wedge }{K_L^{(0)}}(s,\tau )=\int\limits_0^\infty
dxx^{s-1}K_L^{(0)}(x,\tau )= 
\]

\[
\frac 1{2\pi \hbar }\int\limits_0^\infty dx\,x^{s-1}\int\limits_{-\infty
}^\infty dp\cdot \exp \left\{ i\frac{px}\hbar -i\frac{D_\alpha |p|^\alpha
\tau }\hbar \right\} . 
\]

By a change of variables

\[
p\rightarrow \left( \frac \hbar {iD_\alpha \tau }\right) ^{1/\alpha
}\varsigma ,\qquad x\rightarrow \hbar \left( \frac \hbar {iD_\alpha \tau
}\right) ^{-1/\alpha }\xi , 
\]

$\stackrel{\wedge }{K^{(0)}}(s,\tau )$ goes over to

\begin{equation}
\stackrel{\wedge }{K_L^{(0)}}(s,\tau )=\frac 1{2\pi }\left( \frac \hbar
{(\hbar /iD_\alpha \tau )^{1/\alpha }}\right) ^{s-1}\int\limits_0^\infty
d\xi \xi ^{s-1}\int\limits_{-\infty }^\infty d\varsigma \exp \{i\varsigma
\xi -|\varsigma |^\alpha \}.  \label{eq20}
\end{equation}

The integrals over $d\xi $ and $d\varsigma $ can be evaluated by using the
equations (3.3) and (3.6) of the Ref. \cite{West2}. Indeed, we have

\begin{equation}
\int\limits_0^\infty d\xi \xi ^{s-1}\int\limits_0^\infty d\varsigma \exp
\{i\varsigma \xi -\varsigma ^\alpha \}=\frac 4{s-1}\sin \frac{\pi (s-1)}%
2\Gamma (s)\Gamma (1-\frac{s-1}\alpha ),  \label{eq21}
\end{equation}

where $s-1<\alpha \leq 2$ and $\Gamma (s)$ is the gamma function\footnote{%
The gamma function $\Gamma (s)$ has the familiar integral representation $%
\Gamma (s)=\int\limits_0^\infty dtt^{s-1}e^{-t}$, $\mathrm{Res}>0$.}.

Inserting Eq.(\ref{eq21}) into Eq.(\ref{eq20}) and using the functional
relations for the gamma function, $\Gamma (1-z)=-z\Gamma (-z)$ and $\Gamma
(z)\Gamma (1-z)=\pi /\sin \pi z$, we find

\[
\stackrel{\wedge }{K_L^{(0)}}(s,\tau )=\frac 1\alpha \left( \frac \hbar
{(\hbar /iD_\alpha \tau )^{1/\alpha }}\right) ^{s-1}\frac{\Gamma (s)\Gamma (%
\frac{1-s}\alpha )}{\Gamma (\frac{1-s}2)\Gamma (\frac{1+s}2)}. 
\]

Then the inverse Mellin transform gives a free particle quantum mechanical
kernel $K_L^{(0)}(x,\tau )$

\[
K_L^{(0)}(x,\tau )=\frac 1{2\pi i}\int\limits_{c-i\infty }^{c+i\infty
}dsx^{-s}\stackrel{\wedge }{K_L^{(0)}}(s,\tau )= 
\]

\[
\frac 1{2\pi i}\frac 1\alpha \cdot \int\limits_{c-i\infty }^{c+i\infty
}ds\left( \frac \hbar {(\hbar /iD_\alpha \tau )^{1/\alpha }}\right)
^{s-1}x^{-s}\frac{\Gamma (s)\Gamma (\frac{1-s}\alpha )}{\Gamma (\frac{1-s}%
2)\Gamma (\frac{1+s}2)}, 
\]

where the integration path is the straight line from $c-i\infty $ to $%
c+i\infty $ with $0<c<1$. Replacing $s$ by $-s$ we find

\begin{equation}
K_L^{(0)}(x,\tau )=  \label{eq22}
\end{equation}

\[
\frac 1\alpha \left( \frac \hbar {(\hbar /iD_\alpha \tau )^{1/\alpha
}}\right) ^{-1}\frac 1{2\pi i}\int\limits_{-c-i\infty }^{-c+i\infty
}ds\left( \frac 1\hbar \left( \frac \hbar {iD_\alpha \tau }\right)
^{1/\alpha }x\right) ^s\frac{\Gamma (-s)\Gamma (\frac{1+s}\alpha )}{\Gamma (%
\frac{1+s}2)\Gamma (\frac{1-s}2)}. 
\]

The path of integration may be deformed into one running clockwise around $%
R_{+}-c$. Comparison with the definition of the Fox's $H$-function (see Eqs.(%
\ref{eqA.1})-(\ref{eqA.3}), Appendix A) yields

\begin{equation}
K_L^{(0)}(x,\tau )=  \label{eq23}
\end{equation}

\[
\frac 1\alpha \left( \frac \hbar {(\hbar /iD_\alpha \tau )^{1/\alpha
}}\right) ^{-1}H_{2,2}^{1,1}\left[ \frac 1\hbar \left( \frac \hbar
{iD_\alpha \tau }\right) ^{1/\alpha }x\mid \QATOP{(1-1/\alpha ,1/\alpha
),(1/2,1/2)}{(0,1),(1/2,1/2)}\right] . 
\]

Because of the Property 12.2.5 (see Appendix A) of the Fox's $H$-function
the kernel $K_L^{(0)}(x,\tau )$ becomes

\begin{equation}
K_L^{(0)}(x,\tau )=\frac 1{\alpha x}H_{2,2}^{1,1}\left[ \frac 1\hbar \left(
\frac \hbar {iD_\alpha \tau }\right) ^{1/\alpha }x\mid \QATOP{(1,1/\alpha
),(1,1/2)}{(1,1),(1,1/2)}\right] ,\qquad x>0.  \label{eq24}
\end{equation}

Hence, for any $x$ we have,

\begin{equation}
K_L^{(0)}(x,\tau )=\frac 1{\alpha x}H_{2,2}^{1,1}\left[ \frac 1\hbar \left(
\frac \hbar {iD_\alpha \tau }\right) ^{1/\alpha }|x|\mid \QATOP{(1,1/\alpha
),(1,1/2)}{(1,1),(1,1/2)}\right] ,  \label{eq25}
\end{equation}
Remembering Eqs.(\ref{eq16}), (\ref{eq17}) finally yields

\[
K_L^{(0)}(x_bt_b|x_at_a)= 
\]

\begin{equation}
\frac 1{\alpha |x_b-x_a|}H_{2,2}^{1,1}\left[ \frac 1\hbar \left( \frac \hbar
{iD_\alpha (t_b-t_a)}\right) ^{1/\alpha }|x_b-x_a|\mid \QATOP{(1,1/\alpha
),(1,1/2)}{(1,1),(1,1/2)}\right] .  \label{eq26}
\end{equation}

This is new equation for 1-D free particle quantum kernel $%
K_L^{(0)}(x_bt_b|x_at_a)$.

Let us show that Eq.(\ref{eq26}) includes as a particular case at $\alpha =2$
the well known Feynman quantum mechanical kernel given by Eq.(\ref{eq8}).
Indeed, setting in Eq.(\ref{eq26}) $\alpha =2$ and applying the series
expansion Eq.(\ref{eqA.4}) for the function $H_{2,2}^{1,1}[\frac 1\hbar
\left( \frac \hbar {D_2\tau }\right) ^{1/2}|x|\mid \QATOP{(1,1/2),(1,1/2)}{%
(1,1),(1,1/2)}]$ we have

\[
K_L^{(0)}(x,\tau )|_{\alpha =2}=\frac 1{2\hbar }\left( \frac \hbar {iD_2\tau
}\right) ^{1/2}\sum\limits_{k=0}^\infty \left( -\frac 1\hbar \left( \frac
\hbar {iD_2\tau }\right) ^{1/2}\right) ^k\frac{|x|^k}{k!}\frac 1{\Gamma (%
\frac{1-k}2)}. 
\]

By substituting $k\rightarrow 2l$ we obtain

\begin{equation}
K_L^{(0)}(x,\tau )|_{\alpha =2}=\frac 1{2\hbar }\left( \frac \hbar {iD_2\tau
}\right) ^{1/2}\sum\limits_{l=0}^\infty \left( -\frac 1\hbar \left( \frac
\hbar {iD_2\tau }\right) ^{1/2}\right) ^{2l}\frac{x^{2l}}{(2l)!}\frac
1{\Gamma (\frac 12-l)}.  \label{eq27}
\end{equation}

Taking into account the identity

\[
\Gamma (\frac 12+z)\Gamma (\frac 12-z)=\frac \pi {\cos \pi z}, 
\]

and applying the Gauss multiplication formula

\[
\Gamma (2l)=\sqrt{\frac{2^{4l-1}}{2\pi }}\Gamma (l)\Gamma (l+\frac 12), 
\]

we find that

\begin{equation}
(2l)!\Gamma (\frac 12-l)=\frac{\sqrt{\pi }}{(-1)^l}(2)^{2l}l!.  \label{eq28}
\end{equation}

With the help of Eq.(\ref{eq28}) the kernel $K_L^{(0)}(x,\tau )|_{\alpha =2}$
can be rewritten as

\begin{equation}
K_L^{(0)}(x,\tau )|_{\alpha =2}=\frac 1{2\sqrt{\pi }\hbar }\left( \frac
\hbar {iD_2\tau }\right) ^{1/2}\sum\limits_{l=0}^\infty \left( -\frac 1\hbar
\left( \frac \hbar {iD_2\tau }\right) ^{1/2}\right) ^{2l}\frac{(-1)^lx^{2l}}{%
2^{2l}l!}=  \label{eq29}
\end{equation}

\[
\frac 1{2\sqrt{\pi }\hbar }\left( \frac \hbar {iD_2\tau }\right) ^{1/2}\exp
\{-\frac 14\frac{x^2}{\hbar iD_2\tau }\}. 
\]

Since $D_2=1/2m$ \cite{Laskin1}, \cite{Laskin2} we finally obtain for the
Feynman kernel $K^{(0)}(x,\tau )$

\begin{equation}
K^{(0)}(x,\tau )\equiv K_L^{(0)}(x,\tau )|_{\alpha =2}=\sqrt{\frac m{2\pi
i\hbar \tau }}\exp \{\frac{imx^2}{2\hbar \tau }\}.  \label{eq30}
\end{equation}

It turns into Eq.(3-3) of Ref. \cite{Feynman1} if we use Eq.(\ref{eq16}).

Thus, it is shown that a free particle Feynman kernel can be derived from
the general equation (\ref{eq26}).

\subsection{3-D generalization}

The above developments can be generalized to 3-D dimension. It is obviously
that a free particle kernel given by Eq.(\ref{eq10}) for the 3-D case has
the form

\begin{equation}
K_L^{(0)}(\mathbf{r}_bt_b|\mathbf{r}_at_a)=\frac 1{(2\pi \hbar
)^3}\int\limits_{}^{}d^3p\cdot \exp \left\{ i\frac{\mathbf{p}(\mathbf{r}_b-%
\mathbf{r}_a)}\hbar -i\frac{D_\alpha |\mathbf{p}|^\alpha (t_b-t_a)}\hbar
\right\} ,  \label{eq31}
\end{equation}

where $\mathbf{r}$ and $\mathbf{p}$ are the 3-D vectors.

To express the kernel $K_L^{(0)}(\mathbf{r}_bt_b|\mathbf{r}_at_a)$ in the
terms of the Fox's $H$-function we write

\[
K_L^{(0)}(\mathbf{r}_bt_b|\mathbf{r}_at_a)= 
\]

\[
\frac 1{(2\pi \hbar )^3}\int\limits_0^\infty dpp^2\int\limits_{-\pi }^\pi
d\vartheta \int\limits_0^{2\pi }d\varphi \exp \left\{ i\frac{p|\mathbf{r}_b-%
\mathbf{r}_a|\cos \vartheta }\hbar -i\frac{D_\alpha |\mathbf{p}|^\alpha
(t_b-t_a)}\hbar \right\} = 
\]

\[
\frac 1{2\pi ^2\hbar ^2|\mathbf{r}_b-\mathbf{r}_a|}\int\limits_0^\infty
dpp\sin (\frac{p|\mathbf{r}_b-\mathbf{r}_a|}\hbar )\exp \left\{ -i\frac{%
D_\alpha |\mathbf{p}|^\alpha (t_b-t_a)}\hbar \right\} . 
\]

With help of the formula

\begin{equation}
p\sin (\frac{p|\mathbf{r}_b-\mathbf{r}_a|}\hbar )=-\hbar \frac \partial
{\partial |\mathbf{r}_b-\mathbf{r}_a|}\cos (\frac{p|\mathbf{r}_b-\mathbf{r}%
_a|}\hbar ),  \label{eq32}
\end{equation}

$K_L^{(0)}(\mathbf{r}_bt_b|\mathbf{r}_at_a)$ can be written as

\begin{equation}
K_L^{(0)}(\mathbf{r}_bt_b|\mathbf{r}_at_a)=-\frac 1{2\pi }\frac \partial
{\partial x}K_L^{(0)}(x;t_b-t_a)|_{x=|\mathbf{r}_b-\mathbf{r}_a|},
\label{eq33}
\end{equation}

where the kernel $K_L^{(0)}(x;t_b-t_a)$ is 1-D kernel given by Eq.(\ref{eq25}%
).

Let us note that Eq.(\ref{eq33}) is just a special case of a general
relation that holds between the D-mensional and D+2-mensional Fourier
transforms of any isotropic function. An important consequence of Eq.(\ref
{eq33}) is that it allows us to evaluate the 3D kernel based on Eq.(\ref
{eq25}) for 1-D quantum kernel. Thus, the problem is to calculate the
derivative of the 1-D kernel. Using the Property 12.2.9 (see Appendix A)
yields

\begin{equation}
K_L^{(0)}(\mathbf{r}_bt_b|\mathbf{r}_at_a)=  \label{eq34}
\end{equation}

\[
-\frac 1{2\pi \alpha }\frac 1{|\mathbf{r}_b-\mathbf{r}_a|^3}H_{3,3}^{1,2}%
\left[ \frac 1\hbar \left( \frac \hbar {iD_\alpha (t_b-t_a)}\right)
^{1/\alpha }|\mathbf{r}_b-\mathbf{r}_a|\mid \QATOP{(1,1),(1,1/\alpha
),(1,1/2)}{(1,1),(1,1/2),(2,1)}\right] . 
\]

This is new equation for a free particle quantum mechanical 3-D kernel. We
see that in comparison with 1-D case the 3-D quantum kernel is expressed in
the terms of $H_{3,3}^{1,2}$ Fox's $H$-function.

Let's see that the 3-D kernel given by Eq.(\ref{eq34}) goes over into the
Feynman 3-D kernel. Setting in Eq.(\ref{eq34}) $\alpha =2$ and applying the
series expansion Eq.(\ref{eqA.4}) for the function $H_{3,3}^{1,2}$ we have 
\[
K_L^{(0)}(\mathbf{r}_bt_b|\mathbf{r}_at_a)|_{\alpha =2}= 
\]

\[
-\frac 1{4\pi \hbar |\mathbf{r}_b-\mathbf{r}_a|}\sum\limits_{k=0}^\infty 
\frac{\Gamma (1+k)}{\Gamma (k)\Gamma (\frac{1-k}2)}\frac{(-1)^k}{k!}\cdot
\left( \left( \frac \hbar {iD_\alpha (t_b-t_a)}\right) ^{1/2}|\mathbf{r}_b-%
\mathbf{r}_a|\right) ^{1+k}, 
\]

or

\begin{equation}
K_L^{(0)}(\mathbf{r}_bt_b|\mathbf{r}_at_a)|_{\alpha =2}=  \label{eq35}
\end{equation}

\[
-\frac 1{4\pi \hbar }\cdot \left( \frac \hbar {iD_2(t_b-t_a)}\right)
^{1/2}\frac \partial {\partial |\mathbf{r}_b-\mathbf{r}_a|}\sum%
\limits_{k=0}^\infty \frac{\left\{ -\frac 1\hbar \left( \frac \hbar
{iD_2(t_b-t_a)}\right) ^{1/2}|\mathbf{r}_b-\mathbf{r}_a|\right\} ^k}{%
k!\Gamma (\frac{1-k}2)}. 
\]

By substituting $k\rightarrow 2l$ and using Eq.(\ref{eq28}) we observe that
there exists the following relation

\begin{equation}
\sum\limits_{k=0}^\infty \frac{\left\{ -\frac 1\hbar \left( \frac \hbar
{iD_2(t_b-t_a)}\right) ^{1/2}|\mathbf{r}_b-\mathbf{r}_a|\right\} ^k}{%
k!\Gamma (\frac{1-k}2)}=\frac 1{\sqrt{\pi }}\exp \left\{ -\frac{|\mathbf{r}%
_b-\mathbf{r}_a|^2}{4i\hbar D_2(t_b-t_a)}\right\} ,  \label{eq36}
\end{equation}

which transforms Eq.(\ref{eq35}) into

\begin{equation}
K_L^{(0)}(\mathbf{r}_bt_b|\mathbf{r}_at_a)|_{\alpha =2}=\left( \frac \hbar
{4\pi i\hbar D_2(t_b-t_a)}\right) ^{3/2}\exp \left\{ -\frac{|\mathbf{r}_b-%
\mathbf{r}_a|^2}{4i\hbar D_2(t_b-t_a)}\right\} .  \label{eq37}
\end{equation}

Since $D_2=1/2m$ we got the Feynman 3-D quantum mechanical kernel $K^{(0)}(%
\mathbf{r}_bt_b|\mathbf{r}_at_a)$ of a particle with the mass $m$ (see, 
\textit{Problem 4-12}, page 89 of Ref. \cite{Feynman1}),

\[
K^{(0)}(\mathbf{r}_bt_b|\mathbf{r}_at_a)\equiv K_L^{(0)}(\mathbf{r}_bt_b|%
\mathbf{r}_at_a)|_{\alpha =2}=\left( \frac m{2\pi i\hbar (t_b-t_a)}\right)
^{3/2}\exp \left\{ \frac{im|\mathbf{r}_b-\mathbf{r}_a|^2}{2\hbar (t_b-t_a)}%
\right\} . 
\]

Thus, the general equation (\ref{eq34}) includes the Feynman 3-D kernel as a
special case at $\alpha =2$.

\section{Transforms of a free particle kernel}

\subsection{The Laplace transform of a free particle kernel}

The Laplace in time transform $\widetilde{K}_L^{(0)}(x,s)$ of 1-D free
particle kernel is defined as

\begin{equation}
\widetilde{K}_L^{(0)}(x,s)=\int\limits_0^\infty d\tau e^{-s\tau
}K_L^{(0)}(x,\tau ),  \label{eq38}
\end{equation}

where $K_L^{(0)}(x,\tau )$ is given by Eq.(\ref{eq25}).

Using Eq.(\ref{eq38}) and applying the series expansion for the function $%
H_{2,2}^{1,1}$ yield

\begin{equation}
\widetilde{K}_L^{(0)}(x,s)=\frac 1{\alpha x}\sum\limits_{k=0}^\infty \frac{%
\Gamma (\frac{1+k}\alpha )}{\Gamma (\frac{1+k}2)\Gamma (\frac{1-k}2)}\frac{%
(-1)^k}{k!}\cdot \left( \frac 1\hbar \left( \frac \hbar {iD_\alpha }\right)
^{1/\alpha }|x|\right) ^{1+k}\int\limits_0^\infty d\tau e^{-s\tau }\tau ^{-%
\frac{1+k}\alpha }=  \label{eq39}
\end{equation}

\[
\frac 1{\alpha xs}\sum\limits_{k=0}^\infty \frac{\Gamma (\frac{1+k}\alpha
)\Gamma (1-\frac{1+k}\alpha )}{\Gamma (\frac{1+k}2)\Gamma (\frac{1-k}2)}%
\frac{(-1)^k}{k!}\cdot \left( \frac 1\hbar \left( \frac{\hbar s}{iD_\alpha }%
\right) ^{1/\alpha }|x|\right) ^{1+k}, 
\]

where we took into account that

\[
\int\limits_0^\infty d\tau e^{-s\tau }\tau ^{-\frac{1+k}\alpha }=s^{\frac{1+k%
}\alpha -1}\Gamma (1-\frac{1+k}\alpha ). 
\]

Applying the definition of the Fox's function $H_{3,2}^{1,2}$ (see Eqs.(\ref
{eqA.1})-(\ref{eqA.3}), Appendix A) we can write finally the Laplace
transform $\widetilde{K}_L^{(0)}(x,s)$ in terms of $H$-function

\begin{equation}
\widetilde{K}_L^{(0)}(x,s)=\frac 1{\alpha xs}H_{3,2}^{1,2}\left[ \frac
1\hbar \left( \frac{\hbar s}{iD_\alpha }\right) ^{1/\alpha }|x|\mid \QATOP{%
(1,1/\alpha ),(0,-1/\alpha ),(1,1/2)}{(1,1),(1,1/2)}\right] .  \label{eq40}
\end{equation}

Putting in Eq.(\ref{eq40}) $\alpha =2$ and using the series expansion for
the $H_{3,2}^{1,2}$-function we obtain the Laplace transform of a free
particle kernel for the standard quantum mechanics 
\begin{equation}
\widetilde{K}^{(0)}(x,s)\equiv \widetilde{K}_L^{(0)}(x,s)|_{\alpha =2}=\sqrt{%
\frac m{2si\hbar }}\exp \left\{ -\sqrt{\frac{2ms}{i\hbar }}|x|\right\} .
\label{eq41}
\end{equation}

\subsection{The energy-time transformation}

In quantum mechanics an important role plays the Fourier transform of the
kernel in the time variable, which is the fixed-energy kernel $%
k_L(x_2,x_1;E) $

\begin{equation}
k_L(x_2,x_1;E)=\int\limits_{t_1}^\infty dt_2e^{(i/\hbar )E(t_2-t_1)}\cdot
K_L(x_2t_2|x_1t_1),  \label{eq42}
\end{equation}

where $K_L(x_2t_2|x_1t_1)$ is given by Eq.(\ref{eq12}).

To make the integral convergent, we have to move the energy into the upper
complex half-plane by an infinitesimal amount $\epsilon $. Then the L\'evy
fixed-energy kernel becomes

\begin{equation}
k_L(x_2,x_1;E)=\sum\limits_{n=1}^\infty \phi _n(x_2)\phi _n^{*}(x_1)\cdot 
\frac{i\hbar }{E-E_n+i\epsilon },  \label{eq43}
\end{equation}

here $\phi _n(x)$ and $E_n$ are the eigenfunctions and eigenenergies of the
Hamiltonian $H_\alpha $ defined by Eq.(\ref{eq13}).

The small $i\epsilon $-shift in the energy $E$ in (\ref{eq43}) may be
thought of as being attached to each of the energies $E_n$ which are thus
placed by an infinitesimal piece below the real energy axis. When doing the
Fourier integral (\ref{eq42}) the exponential $e^{-(i/\hbar )E(t_2-t_1)}$
makes it always possible to close the integration contour along the energy
axis by an infinite semicircle in the complex energy plane, which lies in
the upper half-plane for $t_2<t_1$ and in the lower half-plane for $t_2>t_1$%
. The $i\epsilon $-shift guarantees that for $t_2<t_1$, there is no pole
inside the closed contour making the kernel vanish. For $t_2>t_1$, on the
other hand, poles in the lower half-plane give, via Cauchy's residue
theorem, the spectral representation of the kernel (\ref{eq43}).

We see that the fixed-energy kernel $k_L(x_2,x_1;E)$ and the kernel $%
K_L(x_2t_2|x_1t_1)$ related each other by the inverse energy Fourier
transform

\begin{equation}
K_L(x_2t_2|x_1t_1)=\frac 1{2\pi \hbar }\int\limits_{-\infty }^\infty
dE\,\,e^{-(i/\hbar )E(t_2-t_1)}\cdot k_L(x_2,x_1;E).  \label{eq44}
\end{equation}

Let us calculate a free particle fixed-energy kernel $k_L^{(0)}(\mathbf{r}_2,%
\mathbf{r}_1;E)$ defined as follows

\begin{equation}
k_L^{(0)}(\mathbf{r}_2,\mathbf{r}_1;E)=\int\limits_{t_1}^\infty
dt_2e^{(i/\hbar )E(t_2-t_1)}\cdot K_L^{(0)}(\mathbf{r}_2t_2|\mathbf{r}_1t_1).
\label{eq45}
\end{equation}

With help of Eq.(\ref{eq34}) we have

\begin{equation}
k_L^{(0)}(\mathbf{r}_2,\mathbf{r}_1;E)=-\frac 1{2\pi \alpha }\frac 1{|%
\mathbf{r}_2-\mathbf{r}_1|^3}\int\limits_{t_1}^\infty dt_2e^{(i/\hbar
)E(t_2-t_1)}\times  \label{eq46}
\end{equation}

\[
H_{3,3}^{1,2}\left[ \frac 1\hbar \left( \frac \hbar {iD_\alpha
(t_2-t_1)}\right) ^{1/\alpha }|\mathbf{r}_2-\mathbf{r}_1|\mid \QATOP{%
(1,1),(1,1/\alpha ),(1,1/2)}{(1,1),(1,1/2),(2,1)}\right] . 
\]

Using the Property 12.2.8 of $H$-function (see Appendix A) yields

\begin{equation}
k_L^{(0)}(\mathbf{r}_2,\mathbf{r}_1;E)=\frac \hbar {2\pi \alpha iE|\mathbf{r}%
_2-\mathbf{r}_1|^3}\times  \label{eq47}
\end{equation}

\[
H_{4,3}^{1,3}\left[ \frac 1\hbar \left( -\frac E{D_\alpha }\right)
^{1/\alpha }|\mathbf{r}_2-\mathbf{r}_1|\mid \QATOP{(0,-1/\alpha
),(1,1),(1,1/\alpha ),(1,1/2)}{(1,1),(1,1/2),(2,1)}\right] . 
\]

Thus the expression for the fixed-energy kernel $k_L^{(0)}(\mathbf{r}_2,%
\mathbf{r}_1;E)$ involves the $H_{4,3}^{1,3}$ Fox's $H$-function. The
inverse energy Fourier transform defined by Eq.(\ref{eq44}) and the
expression (\ref{eq47}) allow to obtain the following alternative
representation for the quantum mechanical kernel $K_L^{(0)}(\mathbf{r}_2t_2|%
\mathbf{r}_1t_1)$

\begin{equation}
K_L^{(0)}(\mathbf{r}_2t_2|\mathbf{r}_1t_1)=\frac \hbar {(2\pi )^2\alpha i|%
\mathbf{r}_2-\mathbf{r}_1|^3}\int\limits_{-\infty }^\infty dE\,\frac{%
e^{-(i/\hbar )E(t_2-t_1)}}E\,\times  \label{eq48}
\end{equation}

\[
H_{4,3}^{1,3}\left[ \frac 1\hbar \left( -\frac E{D_\alpha }\right)
^{1/\alpha }|\mathbf{r}_2-\mathbf{r}_1|\mid \QATOP{(0,-1/\alpha
),(1,1),(1,1/\alpha ),(1,1/2)}{(1,1),(1,1/2),(2,1)}\right] . 
\]

If we put $\alpha =2$ Eq.(\ref{eq48}) goes over into the standard quantum
mechanical fixed-energy kernel $k_L^{(0)}(\mathbf{r}_2,\mathbf{r}%
_1;E)|_{\alpha =2}=k^{(0)}(\mathbf{r}_2,\mathbf{r}_1;E)$. Indeed, setting in
Eq.(\ref{eq48}) $\alpha =2$ and taking into account that in accordance with
the definition of $H$-function (see Eqs.(\ref{eqA.1})-(\ref{eqA.3}),
Appendix A)

\[
H_{4,3}^{1,3}\left[ \kappa \cdot |\mathbf{r}_2-\mathbf{r}_1|\mid \QATOP{%
(0,-1/2),(1,1),(1,1/2),(1,1/2)}{(1,1),(1,1/2),(2,1)}\right] = 
\]

\[
-\left( \kappa \cdot |\mathbf{r}_2-\mathbf{r}_1|\right) ^2\exp \left(
-\kappa \cdot |\mathbf{r}_2-\mathbf{r}_1|\right) , 
\]

we find the well known equation for the fixed-energy kernel (see, for
example Eq.(1.390) at the space dimension $D=3$, Chapter 1, \cite{Kleinert})

\begin{equation}
k^{(0)}(\mathbf{r}_2,\mathbf{r}_1;E)=\frac{2m}{4\pi \hbar i\kappa ^2|\mathbf{%
r}_2-\mathbf{r}_1|^3}\times \left( \kappa \cdot |\mathbf{r}_2-\mathbf{r}%
_1|\right) ^2\exp \left( -\kappa \cdot |\mathbf{r}_2-\mathbf{r}_1|\right) =
\label{eq49}
\end{equation}

\[
\frac m{2\pi \hbar i|\mathbf{r}_2-\mathbf{r}_1|}\times \exp \left( -\kappa
\cdot |\mathbf{r}_2-\mathbf{r}_1|\right) , 
\]

where for simplicity the notation $\kappa =\frac 1\hbar \left( -\frac
E{D_\alpha }\right) ^{1/\alpha }$ has been introduced.

\subsection{Momentum representation}

To find quantum kernel in the momentum representation let us introduce the
momentum space wave function $\varphi (\mathbf{p},t)$, 
\[
\varphi (\mathbf{p},t)=\int d\mathbf{r}e^{-\frac i\hbar \mathbf{pr}}\psi (%
\mathbf{r},t), 
\]

where $\psi (\mathbf{r},t)$ is the wave function in coordinate representation

\[
\psi (\mathbf{r},t)=\frac 1{(2\pi \hbar )^3}\int d\mathbf{p}e^{\frac i\hbar 
\mathbf{pr}}\varphi (\mathbf{p},t). 
\]

Then we consider the 3-D generalization of Eq.(\ref{eq4})

\[
\psi (\mathbf{r}_b,t_b)=\int d\mathbf{r}_aK_L(\mathbf{r}_bt_b|\mathbf{r}%
_at_a)\cdot \psi (\mathbf{r}_a,t_a). 
\]

Substituting the equation for the wave function in coordinate representation 
$\psi (\mathbf{r},t)$ in terms of the wave function in the momentum
representation $\varphi (\mathbf{p},t)$ yields

\[
\varphi (\mathbf{p}_b,t_b)=\int d\mathbf{p}_a\mathcal{K}_L(\mathbf{p}_bt_b|%
\mathbf{p}_at_a)\cdot \varphi (\mathbf{p}_a,t_a), 
\]

where the kernel in the momentum representation $\mathcal{K}_L(\mathbf{p}%
_bt_b|\mathbf{p}_at_a)$ is defined in terms of the kernel in coordinate
representation $K_L(\mathbf{r}_bt_b|\mathbf{r}_at_a)$ as follows

\[
\mathcal{K}_L(\mathbf{p}_bt_b|\mathbf{p}_at_a)=\int d\mathbf{r}_bd\mathbf{r}%
_ae^{-\frac i\hbar \mathbf{p}_b\mathbf{r}_b+\frac i\hbar \mathbf{p}_a\mathbf{%
r}_a}\cdot K_L(\mathbf{r}_bt_b|\mathbf{r}_at_a). 
\]

For example, for a free particle we have

\begin{equation}
\mathcal{K}_L^{(0)}(\mathbf{p}_bt_b|\mathbf{p}_at_a)=\int d\mathbf{r}_bd%
\mathbf{r}_ae^{-\frac i\hbar \mathbf{p}_b\mathbf{r}_b+\frac i\hbar \mathbf{p}%
_a\mathbf{r}_a}\cdot K_L^{(0)}(\mathbf{r}_bt_b|\mathbf{r}_at_a)=
\label{eq50}
\end{equation}

\[
=(2\pi \hbar )^3\delta (\mathbf{p}_a-\mathbf{p}_b)\cdot \exp \{-\frac i\hbar
D_\alpha |\mathbf{p}_a|^\alpha (t_b-t_a)\},\quad \mathrm{for}\quad t_b>t_a. 
\]

We see from Eq.(\ref{eq50}) that a free particle kernel in the momentum
space is expressed in terms of exponential function, while the kernel in the
coordinate space has more complicated form.

\section{Particle in a box}

Now we are going to consider the impact of integration over the L\'evy
flights paths on quantum kernel for a particle in 1-D box of length $2a$
confined by infinitely high walls at $x=-a$ and $x=a$ . From the eigenvalues 
\cite{Laskin2}

\begin{equation}
E_n=D_\alpha \left( \frac{\pi \hbar }a\right) ^\alpha n^\alpha ,\qquad
\qquad 1<\alpha \leq 2,  \label{eq51}
\end{equation}

with the principal quantum number $n=1,2,3....,$ and corresponding
eigenfunctions \cite{Laskin2}

\begin{equation}
\psi _n(x)=\frac 1{\sqrt{a}}\sin \frac{n\pi x}a,  \label{eq52}
\end{equation}

it follows that the quantum kernel for a particle in the box has the form

\begin{equation}
K_{\mathrm{box}}(x_bt|x_a0)=\frac 1a\sum\limits_{n=1}^\infty \sin \frac{n\pi
x_b}a\sin \frac{n\pi x_a}a\exp \{-\frac i\hbar D_\alpha \left( \frac{\pi
\hbar }a\right) ^\alpha n^\alpha t\}.  \label{eq53}
\end{equation}

Here $x_a$ and $x_b$ are initial and final particle positions in the box.
Note that the eigenfunctions $\psi _n(x)$ given by Eq.(\ref{eq52}) guarantee
that the kernel satisfies the boundary conditions

\begin{equation}
K_{\mathrm{box}}(x_b=a,t|x_a,0)=K_{\mathrm{box}}(x_b,t|x_a=-a,0)=0,
\label{eq54}
\end{equation}

enforced by the two infinite walls at $x=-a$ and $x=a$ at all times.

Then Eq.(\ref{eq53}) can be expressed as

\begin{equation}
K_{\mathrm{box}}(x_bt|x_a0)=  \label{eq55}
\end{equation}

\[
\frac 1{2a}\sum\limits_{n=1}^\infty \left\{ \cos \frac{n\pi }a(x_b-x_a)-\cos 
\frac{n\pi }a(x_b+x_a)\right\} \exp \{-\frac i\hbar D_\alpha \left( \frac{%
\pi \hbar }a\right) ^\alpha n^\alpha t\} 
\]

The next steps to transform the above equation are

\[
K_{\mathrm{box}}(x_bt|x_a0)= 
\]

\[
\frac \pi {2\pi \hbar a}\sum\limits_{l=-\infty }^\infty \int\limits_{-\infty
}^\infty dp\delta (\frac p\hbar -\frac \pi al)\left\{ \exp [\frac{ip(x_b-x_a)%
}\hbar ]-\exp [\frac{-ip(x_b+x_a)}\hbar ]\right\} \times 
\]
\[
\exp \{-\frac i\hbar D_\alpha |p|^\alpha t\}= 
\]

\[
\frac 1{2\pi \hbar }\sum\limits_{l=-\infty }^\infty \int\limits_{-\infty
}^\infty dp\left\{ \exp [\frac{ip(x_b-x_a+2la)}\hbar ]-\exp [\frac{%
-ip(x_b+x_a-2la)}\hbar ]\right\} \times 
\]
\[
\exp \{-\frac i\hbar D_\alpha |p|^\alpha t\}. 
\]

where the Poisson summation formula\footnote{%
\[
\frac \pi a\sum\limits_{l=-\infty }^\infty \delta (\frac p\hbar -\frac \pi
al)=\sum\limits_{l=-\infty }^\infty \mathrm{exp}\{i\frac{2pa}\hbar l\}, 
\]
\par
where $\delta $ is the Dirac delta function.} has been applied.

If we take into account the definition of a free particle kernel $%
K_L^{(0)}(x_bt|x_a0)$ given by Eq.(\ref{eq10}) the kernel for the particle
in the box $K_{\mathrm{box}}(x_bt|x_a0)$ becomes

\begin{equation}
K_{\mathrm{box}}(x_bt|x_a0)=\sum\limits_{l=-\infty }^\infty \left\{
K_L^{(0)}(x_b+2la,t|x_a0)-K_L^{(0)}(-x_b+2la,t|x_a0)\right\} .  \label{eq56}
\end{equation}

In terms of Fox's $H$-function $K_{\mathrm{box}}(x_bt|x_a0)$ is

\begin{equation}
K_{\mathrm{box}}(x_bt|x_a0)=\frac 1\alpha \left( \frac \hbar {(\hbar
/iD_\alpha t)^{1/\alpha }}\right) ^{-1}\times  \label{eq57}
\end{equation}

\[
\sum\limits_{l=-\infty }^\infty \{H_{2,2}^{1,1}\left[ \frac 1\hbar \left(
\frac \hbar {iD_\alpha t}\right) ^{1/\alpha }|x_b-x_a+2la|\mid \QATOP{%
(1-1/\alpha ,1/\alpha ),(1/2,1/2)}{(0,1),(1/2,1/2)}\right] - 
\]

\[
H_{2,2}^{1,1}\left[ \frac 1\hbar \left( \frac \hbar {iD_\alpha t}\right)
^{1/\alpha }|-x_b-x_a+2la|\mid \QATOP{(1-1/\alpha ,1/\alpha ),(1/2,1/2)}{%
(0,1),(1/2,1/2)}\right] \}. 
\]

To understand Eq.(\ref{eq57}) for $K_{\mathrm{box}}(x_bt|x_a0)$ from point
of view of physics let us remind the analogy with the method of image
charges in electrostatics. Following the method of image charges one can
account the conducted plate by putting a negative charge complementing the
positive charge. In this case, the image method yields an infinite number of
charges of alternating signs. The original positive charge gives rise to two
negative charges which are each an image corresponding to one of the two
conducted plates. These images generate mirror images corresponding to other
conducted plate and the image generation process has going on up to
infinity. The quantum problem of particle in the box can be treated similar
to the electrostatics problem of a charge between two parallel conducted
plates. To apply electrostatic analogy to the quantum kernel let us start
with simple physical situation of just one infinite wall and take a look at
all paths running between $x_a$ and $x_b$ in time $t$. As can be seen from
the space-time diagram in Fig.1 there is quantum path which does not cross
the wall and which therefore contributes to the path integral. However
between $x_a$ and $x_b$ there exist also path which crosses the wall an even
number of times. Since this path goes through the forbidden region, it
should not contribute to the path integral. To eliminate contribution of
those paths we have to apply a procedure to ensure that only quantum paths
not crossing the wall will be taken into account. The procedure can be
realized as follows. At first we write down a free particle kernel $%
K_L^{(0)}(x_bt|x_a0)$ which disregards the wall. Then we have to subtract
from $K_L^{(0)}(x_bt|x_a0)$ the contribution of all paths which cross the
wall. This can be done by constructing a free particle path integral with
the same classical mechanical action as one which contributes to the $%
K_L^{(0)}(x_bt|x_a0)$. We take the original path up to the last crossing
with the wall and then to continue along the mirror image of the original
path. We thus end up at the mirror image -$x_b$ of the original end point $%
x_b$. Note that a path running from $x_a$ to -$x_b$ necessarily crosses the
wall at least once. Therefore substracting a free particle kernel between $%
x_a$ and -$x_b$ eliminates contribution from all paths which do not remain
in the physical region $x>0$. Finally we conclude that the kernel $K_{%
\mathrm{wall}}(x_bt|x_a0)$ in presence of a wall can be expressed by
substracting a free particle kernel going from $x_a$ to the mirror image -$%
x_b$ from a free particle kernel going from $x_a$ to $x_b$, see Fig.1,

\[
K_{\mathrm{wall}}(x_bt|x_a0)=K_L^{(0)}(x_bt|x_a0)-K_L^{(0)}(-x_bt|x_a0). 
\]

Expressing the kernel of a particle in the box in terms of free kernel works
exactly in the same way. Indeed, a path intersecting both walls is
subtracted twice, i.e. one time too often. Therefore, one contribution has
to be restored which is done by adding another end point. Continuing the
procedure one ends up with an infinite number of end points. The general
rule to attribute a sign to each end point is that each reflection at a wall
leads to factor -1. This procedure immediately gives us Eq.(\ref{eq57}).

Using the way of consideration which leads from Eq.(\ref{eq17}) to Eq.(\ref
{eq30}) we can see that at $\alpha =2$ the L\'evy quantum kernel Eq.(\ref
{eq57}) goed over into the well known Feynman kernel for a particle in a box
(see, for example Eq.(6.19) and (6.20) in the \cite{Kleinert}).

\section{Conclusion}

We have studied an impact of integration over the paths of the L\'evy
flights on the structure of a free particle quantum mechanical kernel.
Analytical expression of a free particle 1-D quantum kernel has been
obtained in term of the Fox $H$-function. The 3-D generalization has been
presented as well. The Laplace, energy-time and momentum transforms of a
free particle kernel have been obtained and discussed. We have found the
quantum kernel for a particle in the box.

While the Feynman path integral is in fact integration over the
Brownian-like paths the L\'evy path integral is the integral over the L\'evy
flights trajectories. The path integral over the L\'evy flights generalizes
at $\alpha <2$ and becomes at $\alpha =2$ the Feynman path integral. The new
equations (\ref{eq26}), (\ref{eq34}), (\ref{eq40}), (\ref{eq48}), (\ref{eq50}%
) and (\ref{eq57}) at the special case $\alpha =2$ go over into the well
known quantum mechanical equations for a free particle kernel and a particle
in the box kernel.

\section{Appendix A}

\subsection{Fox $H$-function}

In Sec.3 we have expressed a free particle kernel in the term of Fox
function $H_{2,2}^{1,1}$. Apart from the quiet natural way in which the Fox $%
H$-function enters in the path integrals over the L\'evy flights, its
fractional derivatives and integrals are easily calculated by formally
manipulating the parameters in the $H$-function. That is why the Fox's $H$%
-function is important and useful for the L\'evy path integral calculations.

Some properties of the $H$-function in connection with Mellin-Barnes
integrals were investigated by Barnes \cite{Barnes}, Mellin \cite{Mellin},
Dixon and Ferrar \cite{Dixon}. In an attempt to unify and extend the
existing results on symmetrical Fourier kernels, Fox \cite{Fox} has defined
the $H$-function in terms of a general Mellin-Barnes type integral.
Asymptotic expansions and analytic continuations of the Fox function and its
special cases were derived by Braaksma \cite{Braaksma}. Many properties of
the $H$-function are reported in the book \cite{Mathai} along with
applications to statistics.

Fox's $H$-function is defined by the Mellin-Barnes type integral \cite{Fox}, 
\cite{Braaksma} (we follow the notations of the book \cite{Mathai})

\begin{equation}
H_{p,q}^{m,n}(z)=H_{p,q}^{m,n}\left[ z|\QATOP{(a_p,A_p)}{(b_q,B_q)}\right] =
\label{eqA.1}
\end{equation}

\[
H_{p,q}^{m,n}\left[ z\mid \QATOP{(a_1,A_1),...,(a_p,A_p)}{%
(b_1,B_1),...,(b_q,B_q)}\right] =\frac 1{2\pi i}\int\limits_Lds~z^s~\chi
(s), 
\]

where function $\chi (s)$ is given by

\begin{equation}
\chi (s)=\frac{\prod\limits_{j=1}^m\Gamma
(b_j-B_js)\prod\limits_{j=1}^n\Gamma (1-a_j+A_js)}{\prod\limits_{j=m+1}^q%
\Gamma (1-b_j+B_js)\prod\limits_{j=n+1}^p\Gamma (a_j-A_js)},  \label{eqA.2}
\end{equation}

and

\[
z^s=\exp \{s\mathrm{Log}|z|+i\arg z\}, 
\]
here $m$, $n$ $p$ and $q$ are non negative integers satisfying $0\leq n\leq
p $, 1$\leq m\leq q$; and the empty products are interpreted as unity. The
parameters $A_j$ ($j=1,...,p$) and $B_j$ ($j=1,...,q$) are positive numbers; 
$a_j$ ($j=1,...,p$) and $b_j$ ($j=1,...,p$) are complex numbers such that

\begin{equation}
A_j(b_h+\nu )\neq B_h(a_j-\lambda -1),  \label{eqA.3}
\end{equation}
for $\nu $, $\lambda =0,1,...$; $h=1,...,m$; $j=1,...,n$.

The $L$ is a contour separating the points

\[
s=\left( \frac{b_j+\nu }{B_j}\right) ,\quad (j=1,...,m;~~\nu =0,1,...), 
\]

which are the poles of $\Gamma (b_j-B_js)$ $(j=1,...,m)$, from the points

\[
s=\left( \frac{a_j-\nu -1}{A_j}\right) ,\quad (j=1,...,n;~~\nu =0,1,...), 
\]

which are the poles of $\Gamma (1-a_j+A_js)$ $(j=1,...,n)$. The contor $L$
exists on account of (\ref{eqA.3}). These assumptions will be retained
throughout.

In the contracted form the $H$ function in Eq.(\ref{eqA.1}) will be denoted
by one of the following notations:

\[
H(z),\quad H_{p,q}^{m,n}(z),\quad H_{p,q}^{m,n}\left[ z|\QATOP{(a_p,A_p)}{%
(b_q,B_q)}\right] . 
\]

The Fox $H$-function is an analytic function of $z$ which makes sense (i)
for every $z\neq 0$ if $\mu >0$ and (ii) for $0<|z|<\beta ^{-1}$ if $\mu =0$%
, where

\begin{equation}
\mu =\sum\limits_{j=1}^qB_j-\sum\limits_{j=1}^pA_j  \label{eqA.31}
\end{equation}

and

\begin{equation}
\beta =\prod\limits_{j=1}^pA_j^{A_j}\prod\limits_{j=1}^qB_j^{-B_j}.
\label{eqA.32}
\end{equation}

Due to the occurrence of the factor $z^s$ in Eq.(\ref{eqA.1}), the $H$%
-function is in general multiple-valued, but is one-valued on the Riemann
surface of $\log z$.

The $H$ function is a generalization of Meijer's $G$ function \cite{Meijer},
which is also defined by a Mellin-Barnes integral. The $H$ function reduces
to the $G$ function if $A_j=1$ and $B_k=1$ for all $j=1,2,...,p$ and $%
k=1,2,...,q$,

\[
G_{p,q}^{m,n}(z)=H_{p,q}^{m,n}\left[ z\mid \QATOP{(a_1,1),...,(a_p,1)}{%
(b_1,1),...,(b_q,1)}\right] . 
\]

If further $m=1$ and $p\leq q$, then the $H$ function is expressible by

\[
H_{p,q}^{1,n}\left[ z\mid \QATOP{(a_1,1)...(a_p,1)}{(b_1,1)...(b_q,1)}%
\right] =\frac{\prod\limits_{j=1}^n\Gamma (1+b_1-a_j)~z^{b_1}}{%
\prod\limits_{j=2}^q\Gamma (1+b_1-b_j)\prod\limits_{j=n+1}^p\Gamma (a_j-b_1)}%
\times 
\]

\[
\times _pF_{q-1}\left( \QATOP{1+b_1-a_1,...,1+b_1-a_p}{%
1+b_1-b_2,...,1+b_1-b_q};(-1)^{p-n-1}z\right) , 
\]

in terms of generalized hypergeometric functions $_pF_q$ \cite{Mathai}. As
far as many well-known special functions, such as error function, Bessel
functions, Whittaker functions, Jacobi polynomials, and elliptic functions
are included in the class of generalized hypergeometric functions, all of
them can be expressed in the term of the Fox's $H$ function.

To represent of $H$-unction in computable form let us consider the case when
the poles $s=(b_j+\nu )/B_j$ ($j=1,...,m;$ $\nu =0,1,...$) of $%
\prod\limits_{j=1}^m{}^{^{\prime }}\Gamma (b_j-B_js)$ are simple, that is,
where

\[
B_h(b_j+\lambda )\neq B_j(b_h+\nu ),\qquad j\neq h, 
\]

\[
h=1,...,m;\quad \quad \quad \quad \nu ,\lambda =0,1,2,..., 
\]

and the prime means the product without the factor $j=h$. Then we obtain the
following expansion for the $H$-function

\begin{equation}
H_{p,q}^{m,n}(z)=\sum\limits_{h=1}^m\sum\limits_{k=0}^\infty \frac{%
\prod\limits_{j=1}^n\Gamma (1-a_j+A_js_{hk})\prod\limits_{j=1}^m{}^{^{\prime
}}\Gamma (b_j-B_js_{hk})}{\prod\limits_{j=m+1}^q\Gamma
(1-b_j+B_js_{hk})\prod\limits_{j=n+1}^p\Gamma (a_j-A_js_{hk})}\frac{(-1)^k}{%
k!}\frac{z^{s_{hk}}}{B_h},  \label{eqA.4}
\end{equation}

\[
s_{hk}=(b_h+k)/B_h, 
\]

which exists for all $z\neq 0$ if $\mu >0$ and for $0<|z|,\beta ^{-1}$ if $%
\mu =0$, where $\mu $ and $\beta $ are given by Eqs.(\ref{eqA.31}) and (\ref
{eqA.32}).

The formula (\ref{eqA.4}) can be used to calculate the special values of the 
$H$-function and to derive the asymptotic behavior for $z\rightarrow 0$.

\subsection{Some identities of the $H$-function}

The $H$-function possesses many interesting properties that are helpful for
the calculations with the path integrals over L\'evy flights. We present the
list of mainly used properties of the $H$-function. The results of this
section follow readily from the definition of the $H$-function given by Eq.(%
\ref{eqA.1}) and hence no proofs are given here.

\textbf{Property 12.2.1} The $H$-function is symmetric in the pairs $%
(a_1,A_1),...,(a_n,A_n)$, likewise $(a_{n+1},A_{n+1}),...,(a_p,A_p)$; in $%
(b_1,B_1),...,(b_n,B_n)$ and in $(b_{m+1},B_{m+1}),...,(b_q,B_q)$.

\textbf{Property 12.2.2} If one of the $(a_j,A_j)$ $(j=1,...,n)$ is equal to
one of the $(b_j,B_j)$ $(j=m+1,...,q)$ or one of the $(b_j,B_j)$ $%
(j=1,...,m) $ is equal to one of the $(a_j,A_j)$ $(j=n+1,...,p)$], then the $%
H$-function reduces to one of the lower order, and $p,q$ and $n$ (or$)$ $m$
decrease by unity.

Thus, we have the following reduction formula:

\begin{equation}
H_{p,q}^{m,n}\left[ z\mid \QATOP{(a_1,A_1),...,(a_p,A_p)}{%
(b_1,B_1),...,(b_{q-1},B_{q-1}),(a_1,A_1)}\right] =  \label{eqA.5}
\end{equation}

\[
H_{p-1,q-1}^{m,n-1}\left[ z\mid \QATOP{(a_2,A_2),...,(a_p,A_p)}{%
(b_1,B_1),...,(b_{q-1},B_{q-1})}\right] , 
\]

provided $n\geq 1$ and $q>m$.

\textbf{Property 12.2.3}

\begin{equation}
H_{p,q}^{m,n}\left( z\mid \QATOP{(a_1,A_1),...,(a_p,A_p)}{%
(b_1,B_1),...,(b_q,B_q)}\right) =  \label{eqA.6}
\end{equation}

\[
H_{q,p}^{n,m}\left( \frac 1z\mid \QATOP{(1-b_1,B_1),...,(1-b_q,B_q)}{%
(1-a_1,A_1),...,(1-a_p,A_p)}\right) , 
\]

This is an important property of the $H$-function because it enables us to
transform an $H$-function with $\mu
=\sum\limits_{j=1}^mB_j-\sum\limits_{j=1}^nA_j>0$ and $\arg x$ to one with $%
\mu <0$ and $\arg (1/x)$ and vice versa.

\textbf{Property 12.2.4}

\begin{equation}
\frac 1kH_{p,q}^{m,n}\left( z\mid \QATOP{(a_1,A_1),...,(a_p,A_p)}{%
(b_1,B_1),...,(b_q,B_q)}\right) =H_{p,q}^{m,n}\left( z^k\mid \QATOP{%
(a_1,kA_1),...,(a_p,kA_p)}{(b_1,kB_1),...,(b_q,kB_q)}\right) ,  \label{eqA.7}
\end{equation}

where $k>0$.

\textbf{Property 12.2.5}

\begin{equation}
z^\sigma H_{p,q}^{m,n}\left( z\mid \QATOP{(a_1,A_1),...,(a_p,A_p)}{%
(b_1,B_1),...,(b_q,B_q)}\right) =  \label{eqA.8}
\end{equation}

\[
H_{p,q}^{m,n}\left( z\mid \QATOP{(a_1+\sigma A_1,A_1),...,(a_p+\sigma
A_p,A_p)}{(b_1+\sigma B_1,B_1),...,(b_q+\sigma B_q,B_q)}\right) , 
\]

\textbf{Property 12.2.6}

\begin{equation}
H_{p+1,q+1}^{m,n+1}\left( z\mid \QATOP{(0,\gamma ),...,(a_p,A_p)}{%
(b_q,B_q),...,(r,\gamma )}\right) =(-1)^rH_{p+1,q+1}^{m+1,n}\left( z\mid 
\QATOP{(a_p,A_p),...,(0,\gamma )}{(r,\gamma ),...,(b_q,B_q)}\right) ,
\label{eqA.9}
\end{equation}

where $p\leq q$.

\textbf{Property 12.2.7}

\begin{equation}
H_{p+1,q+1}^{m+1,n}\left( z\mid \QATOP{(a_p,A_p),...,(1-r,\gamma )}{%
(1,\gamma ),...,(b_q,B_q)}\right) =  \label{eqA.10}
\end{equation}

\[
(-1)^rH_{p+1,q+1}^{m+1,n}\left( z\mid \QATOP{(1-r,\gamma ),...,(a_p,A_p)}{%
(b_q,B_q),...,(1,\gamma )}\right) , 
\]

where $p\leq q$.

In the above Properties (12.1.2) - (12.1.6) the branches of the $H$-function
are suitably chosen.

\textbf{Property 12.2.8}

\begin{equation}
\int dxx^{\alpha -1}e^{-\sigma x}H_{p,q}^{m,n}\left( \omega x^r\mid \QATOP{%
(a_1,A_1),...,(a_p,A_p)}{(b_1,B_1),...,(b_q,B_q)}\right) =  \label{eqA.11}
\end{equation}

\[
\sigma ^{-\alpha }H_{p+1,q}^{m,n+1}\left( \frac \omega {\sigma ^r}\mid 
\QATOP{(1-\alpha ,r),(a_1,A_1),...,(a_p,A_p)}{(b_1,B_1),...,(b_q,B_q)}%
\right) , 
\]

where $m\cdot n\neq 0$; $r,p\geq 0$; $a,a^{*}>0$; $|\arg \omega |<a^{*}\pi
/2 $; $\mathrm{Re\,}\alpha +r$, $\stackunder{1\leq j\leq m}{\min }\mathrm{Re}%
(b_j/B_j)>0$.

\textbf{Property 12.2.9 }

\begin{equation}
_0D_z^\nu \left[ z^\alpha H_{p,q}^{m,n}\left( (az)^\beta \mid \QATOP{%
(a_j,A_j)}{(b_j,B_j)}\right) \right] =  \label{eqA.12}
\end{equation}

\[
=z^{\alpha -\nu }H_{p+1,q+1}^{m,n+1}\left( (az)^\beta \mid \QATOP{(-\alpha
,\beta ),(a_j,A_j)}{(b_j,B_j),(\nu -\alpha ,\beta )}\right) , 
\]
where $\nu $ is arbitrary, $a$, $b>0$ and $\alpha +\beta \min (b_j/B_j)>-1$ (%
$1\leq j\leq m$). Here the $_0D_z^\nu $ notes fractional derivative of order 
$\nu $ (see for definition Refs. \cite{Oldham}- \cite{Samko}).

\section{Figure caption}

Fig.1. A path crossing the wall is cancelled by a path running to the mirror
point of the end point \cite{Ingold}.

\end{document}